\documentclass[preprint,12pt]{elsarticle}

\newcommand{\Slash}[1]{{\ooalign{\hfil$#1$\hfil\crcr\raise.167ex\hbox{/}}}}

\usepackage{graphicx}
\usepackage{wrapft}
\usepackage{lscape}

\journal{Nuclear Physics A}
\begin{document}

\begin{frontmatter}

\title{Spin-2 $N\Omega$ Dibaryon from Lattice QCD}

\author[label1,label2,label3]{Faisal Etminan}

\author[label1]{Hidekatsu Nemura}

\author[label1,label3]{Sinya Aoki}

\author[label4]{Takumi Doi}

\author[label4,label5]{Tetsuo Hatsuda}

\author[label4]{Yoichi Ikeda}

\author[label6]{Takashi Inoue}

\author[label1]{Noriyoshi Ishii}

\author[label3]{Keiko Murano}

\author[label1]{Kenji Sasaki}

\author{(HAL QCD Collaboration)}

\address[label1]{Center for Computational Sciences, University of Tsukuba, Ibaraki 305-8571, Japan}

\address[label2]{Department of Physics, Faculty of Sciences, University of Birjand ,Birjand 97175-615, Iran}

\address[label3]{Yukawa Institute for Theoretical Physics, Kyoto University, Kyoto 606-8502, Japan}

\address[label4]{Theoretical Research Division, Nishina Center, RIKEN, Saitama 351-0198, Japan}

\address[label5]{Kavli IPMU (WPI), The University of Tokyo, Chiba 277-8583, Japan}

\address[label6]{Nihon University, College of Bioresource Sciences, Kanagawa 252-0880, Japan}

\begin{abstract}
We investigate properties of the $N$(nucleon)-$\Omega$(Omega) interaction in lattice QCD
to seek for possible dibaryon states in the strangeness $-3$ channel.
We calculate the $N \Omega$ potential through the equal-time Nambu-Bethe-Salpeter wave function in 2+1 flavor lattice QCD with the renormalization group improved Iwasaki gauge action and the nonperturbatively $\mathcal{O}(a)$ improved Wilson quark action
at the  lattice spacing $a\simeq 0.12$ fm on a (1.9 fm)$^{3}\times$ 3.8 fm lattice. 
The $ud$ and $s$ quark masses in our study correspond to $m_{\pi}= 875(1)$ MeV and $m_{K}= 916(1)$ MeV.
At these parameter values, the central potential in the S-wave with the spin 2
shows attractions at all distances.
By solving the Schr\"{o}dinger equation with this potential, we find one bound state whose binding energy is 
 $18.9(5.0)(^{+12.1}_{-1.8})$ MeV, where the first error is the statistical one, while the second represents the systematic error.
\end{abstract}

\begin{keyword}
%% keywords here, in the form: keyword \sep keyword

%% MSC codes here, in the form: \MSC code \sep code
%% or \MSC[2008] code \sep code (2000 is the default)

$N\Omega$ interaction\sep $N\Omega$ potential \sep $N\Omega$ dibaryon \sep Lattice QCD

\end{keyword}

\end{frontmatter}

\section{Introduction}

 Possible existence of strange dibaryons 
 is one of the  long standing problems in hadron physics. 
 Among others, the $H$ dibaryon $(uuddss)$ 
 with $J^P=0^+$ and  $I$(isospin)=0~\cite{Jaffe1977},
 and the $N\Omega$ dibaryon ($uudsss$ or $uddsss$) with 
 $J^P=2^+$ and $I$=1/2~\cite{Goldman:1987ma}, are the most interesting candidates.
 Since the Pauli exclusion principle does not operate among quarks 
 in these systems, 
  they can in principle be compact six-quark states unlike 
   the deuteron~\cite{Oka:1988yq,Gal:2010eg}.
 
 Experimentally, if the mass of the $H$ dibaryon is close to 
 $2m_{\Lambda}=2231$ MeV, its 
  strong-decay width is suppressed, so that it may appear as a 
   sharp resonance.  Similarly, if the spin-2 $N\Omega$ dibaryon  is a S-wave
  bound state of $N$ and $\Omega$,   
     the strong decays to octet-decuplet systems  are
  prohibited kinematically \footnote{Note that the ordering of the thresholds in 
    the  octet-decuplet and decuplet-decuplet
  systems with strangeness=$-3$ and charge=$0$  reads
    $P\Omega^-(2611 {\rm MeV}) < \Lambda \Xi^{0*}(2648 {\rm MeV}) 
    < \Xi^0\Sigma^{0*}(2699 {\rm MeV}) < \Xi^{-}\Sigma^{+*} (2705 {\rm MeV})
    < \Sigma^0\Xi^{0*} (2724 {\rm MeV}) \sim  \Sigma^{+}\Xi^{-*} (2724 {\rm MeV}) 
    < \Sigma^{0*}\Xi^{0*} (2916 {\rm MeV}) <  \Sigma^{+*}\Xi^{-*} (2918 {\rm MeV})$.}  
   and those into  octet-octet systems (e.g. $\Lambda\Xi$)
  are  suppressed dynamically due to the D-wave  nature.   
 Therefore, the spin-2 $N\Omega$ dibaryon, if it is a bound state or a sharp resonance,
   could be observed e.g. by relativistic heavy-ion collisions at 
  RHIC and LHC, or by the hadron beam experiments at J-PARC and FAIR.
     
 Although numerous  attempts have been done
 to estimate the binding energy of  strange dibaryon states in QCD motivated models,
  results were  highly dependent on the model assumptions.
 Only recently, first principle calculations based on lattice QCD 
 became available
 for multi hadrons due to the development of  advanced techniques  such as the 
  L\"{u}scher's method \cite{Luscher:1990ux} and the HAL QCD 
  method~\cite{Ishii:2006ec,Aoki:2008hh,Aoki:2009ji,HALQCD:2012aa,HAL2012}.
  In particular,  the $H$-dibaryon from the point of view of baryon-baryon 
 interactions  was  first studied  by  3-flavor QCD simulations in \cite{Inoue:2010hs}, and 
  a possible shallow bound state in the flavor-SU(3) limit was suggested.  
  Such a possibility was later explored and confirmed by more extensive simulations
   with relatively heavy quark masses 
  in  2+1 flavor QCD \cite{Beane:2010hg}  and in  3 flavor QCD  \cite{Inoue:2010es,Inoue:2011ai}. 
 The fate of $H$  in the real world, however, is still uncertain 
 \cite{Beane:2011zpa,Inoue:2011ai,Shanahan:2011su} and needs to be investigated further.

   In this paper, by using  (2+1)-flavor
   lattice QCD simulations with the HAL QCD method  (reviewed in ~\cite{HAL2012}),
    we carry out an exploratory
  study of the spin-2 $N\Omega$ dibaryon. Our strategy is to derive a potential between 
   $N$ and $\Omega$ in the  $^5 S_2$ channel\footnote{We use the standard notation $^{2S+1}L_J$ in the continuous space to specify quantum numbers.} (S-wave and spin-2)
   from the Nambu-Bethe-Salpeter (NBS) wave function measured on the lattice.
   Although the potential itself is not a direct physical observable, it is a useful
    tool to derive various physical quantities such as the
     binding energy and the phase shift  (see the review \cite{HAL2012} for details).
  For relatively heavy $u$ and $d$ quark masses corresponding to
  $m_{\pi}=875$ MeV and  $m_{K}=916$ MeV,
 we find that the $N\Omega$ system has one bound state
  with the binding energy ($B_{N\Omega}$) of about 19 MeV with the statistical error of 5 MeV.    
   Since $B_{N\Omega}$ obtained in quark models  ranges 
  from  negative value (resonance) to  $\mathcal{O}$(100) MeV (deeply bound state)
   \cite{Wang:1995bg,Li:1999bc},
   our exploratory investigations would give useful information
   to both theoretical and experimental studies on the $N\Omega$ dibaryon.

\section{Basic formulation}
\label{sec:formula}

We define the $N\Omega$ potential at low-energies through
the equal-time NBS wave function $\phi(\vec{r})$ 
which satisfies the Schr\"{o}dinger-type equation at low energies,
\begin{equation}
-\frac{1}{2\mu}\nabla^{2}\phi(\vec{r})+\int U(\vec{r},\vec{r^{\prime}})\phi(\vec{r^{\prime}})d^{3}\vec{r^{\prime}}=E\phi(\vec{r}).
\label{eq:potential}
\end{equation}
Here $\mu=m_{N}m_{\Omega}/(m_{N}+m_{\Omega})$  is the
 reduced mass of the $N\Omega$ system, and $E\equiv k^{2}/(2\mu)$
 is the kinetic energy in the center-of-mass frame.
   It is important to note that  the nonlocal potential
$U(\vec{r},\vec{r^{\prime}})$ does not depend on $k$ \cite{Aoki:2009ji}.
At low energies, we 
expand the nonlocal potential in terms of the relative velocity  as\cite{TW67}
$
U(\vec{r},\vec{r^{\prime}})=V_{N\Omega}(\vec{r},\vec{\nabla})\delta(\vec{r}-\vec{r^{\prime}}).
$

The equal-time NBS wave function in the S-wave is obtained  as
\begin{eqnarray}
\phi(r)&=&\frac{1}{24 L^3}{\displaystyle \sum_{\mathcal{R}\in O}} \sum_{\vec{x}}P_{\alpha\beta,\ell}^{S}
%\nonumber \\&\times&
\left\langle 0\left| N_{\alpha}(\mathcal{R} \left[\vec{r}\right]+\vec{x}, t)\Omega_{\beta,\ell}(\vec{x},t)\right| N\Omega;W\right\rangle ,
\\
N_{\alpha}(x)&=&\varepsilon_{abc}(u^{a}(x)C\gamma_{5}d^{b}(x))q^c_{\alpha}(x), \quad
\Omega_{\beta,\ell}(x) =\varepsilon_{abc}s^{a}_\beta(x)(s^{b}(x)C\gamma_{l}s^{c}(x)),\nonumber
\end{eqnarray}
where $W=\sqrt{k^2+m_N^2}+\sqrt{k^2+m_\Omega^2}$ is the total energy,
$\alpha$ and $\beta$ denote Dirac indices, $l$ is a spatial vector index of the gamma matrix $\gamma_l$,
the color indices are given by $a,b,c$ , and $C=\gamma_{4}\gamma_{2}$ is the charge conjugation matrix. 
 Here $N$ corresponds to a proton (neutron) for $q=u (d)$.
The summation over $\mathcal{R} \in O$ is
taken for all cubic-group elements to project out  the $S$-wave
\footnote{Due to the periodic boundary condition, this projection cannot remove
the higher orbital components with $L\geq4$, which however are expected
to be small at low energies.}.
The summation over $\vec{x}$ is taken to project out the state with 
zero total-momentum. 
We take local interpolating  operators $N_{\alpha}(x)$
and $\Omega_{\beta,\ell}(y)$ for the nucleon and $\Omega$:
Although the 
potential depends on the choice of these operators, 
observables do not depend on the choice \cite{Aoki:2009ji}.
Projection operators $(P_{\alpha\beta,\ell}^{S=2})$ are used to select
the spin $2$ state.
It is important to note that the NBS wave function
 at asymptotically large $\vert \vec r\vert$ 
  carries full information of the phases of the S-matrix
    \cite{Lin:2001ek,Aoki:2005uf,Ishizuka2009a, Aoki:2013cra}.

On the lattice, the NBS wave function
 is extracted from the 4-point function as
\begin{eqnarray}
 &  & F_{N\Omega}(\vec{x}-\vec{y},t-t_{0})=\left\langle 0\left| N_{\alpha}(\vec{x},t)\Omega_{\beta,\ell}(\vec{y},t)\overline{J}_{N\Omega}(t_{0})\right|0\right\rangle \nonumber \\
 &  & \simeq\sum_{n}A_{n}\left\langle 0\left| N_{\alpha}(\vec{x},0)\Omega_{\beta,\ell}(\vec{y},0)\right|N\Omega; W_{n}\right\rangle e^{-W_{n}(t-t_{0})}, \\
&& \qquad  A_{n}=\left\langle N\Omega; W_{n}\left|\overline{J}_{N\Omega}(t=0)\right|0\right\rangle \nonumber
\end{eqnarray}
where  $\overline{J}_{N\Omega}(t_{0})$ is a wall-source operator located at
$t_{0}$, 
and  $\left| N\Omega;W_{n}\right\rangle $ is an eigenstate with an eigen-energy $W_{n}$. 
For   $t-t_{0}\gg1$, $F_{N\Omega}$ is dominated by the lowest energy eigenstate.

In the present paper, we consider the single channel scattering in the S-wave and 
 consider the effective central potential  obtained as a leading order term 
 of the velocity expansion:
%\begin{equation}
$V_{C}(r)=\frac{1}{2\mu}\frac{{\nabla}^2\phi(r)}{\phi(r)} + E$.
%\end{equation}
 As shown in  Ref.~\cite{HALQCD:2012aa}, such a potential  can be
  obtained most efficiently by  the time-dependent HAL QCD method,  which does not require
  the difficult task to separate each scattering state:  
\begin{equation}
V_C(r) \simeq \frac{1}{2\mu}{\nabla^2 R(r, t)}/{R(r,t)} -\frac{\partial}{\partial t} \log R(r,t),
\label{eq:t-dep}
\end{equation}
where $R(r,t) = F_{N\Omega}(r, t)/e^{-(m_N+m_\Omega)t}$ with $F_{N\Omega}(r,t)$ 
being obtained from  $F_{N\Omega}(\vec r,t)$ by the S-wave projection. 
In the above equation, we have assumed that relativistic 
corrections $\left( \frac{\partial_t^2}{m_{N,\Omega}}\right)\left( \frac{\partial_t}{m_{N,\Omega}}\right)^{n}$ with $n \ge 0$ are small.  
(For the $NN$ system in the elastic region, such corrections 
 have been  shown to be very small~\cite{HALQCD:2012aa}.)

\section{Numerical results}
\label{sec:result}
\begin{figure}[bth]
\begin{center}
\includegraphics[scale=1]{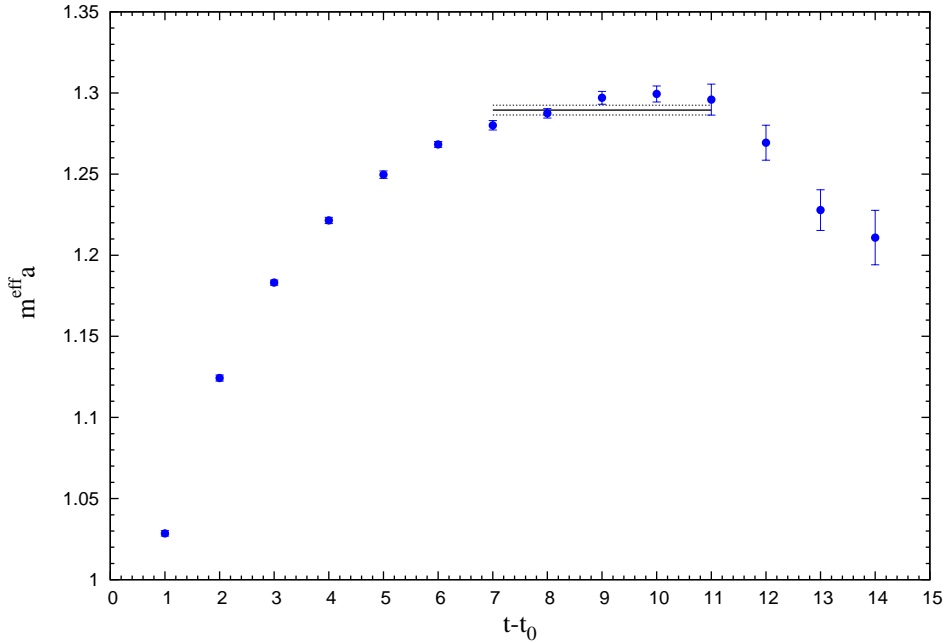}
\caption{Effective mass of  $\Omega$ as a function of $t-t_0$. 
The solid line represents the central value 
in the fit range $t-t_0=7-11$,
while the dashed lines show the statistical uncertainty of the effective mass. }
\label{fig:emass}
\end{center}
\end{figure}

\if0
%
\begin{table}[hbt]
 \caption{Hadron masses in MeV.}
 \label{tab:mass}
 \begin{center}
  \begin{tabular}{|cc|ccccc|}
    \hline
     $m_{\pi}$   & $m_{K}$   & $m_N$   & $m_{\Lambda}$ 
   & $m_{\Sigma}$   & $m_{\Xi}$    & $m_{\Omega}$  \\
    \hline
     875(1) & 916(1) & 1810(2) & 1839(2) & 1846(2) & 1872(2) & 2105(5)   \\
    \hline
  \end{tabular}
 \end{center}
\end{table}
\fi

%---------------------
\begin{figure}[tbh]
\begin{center}
\includegraphics[scale=1]{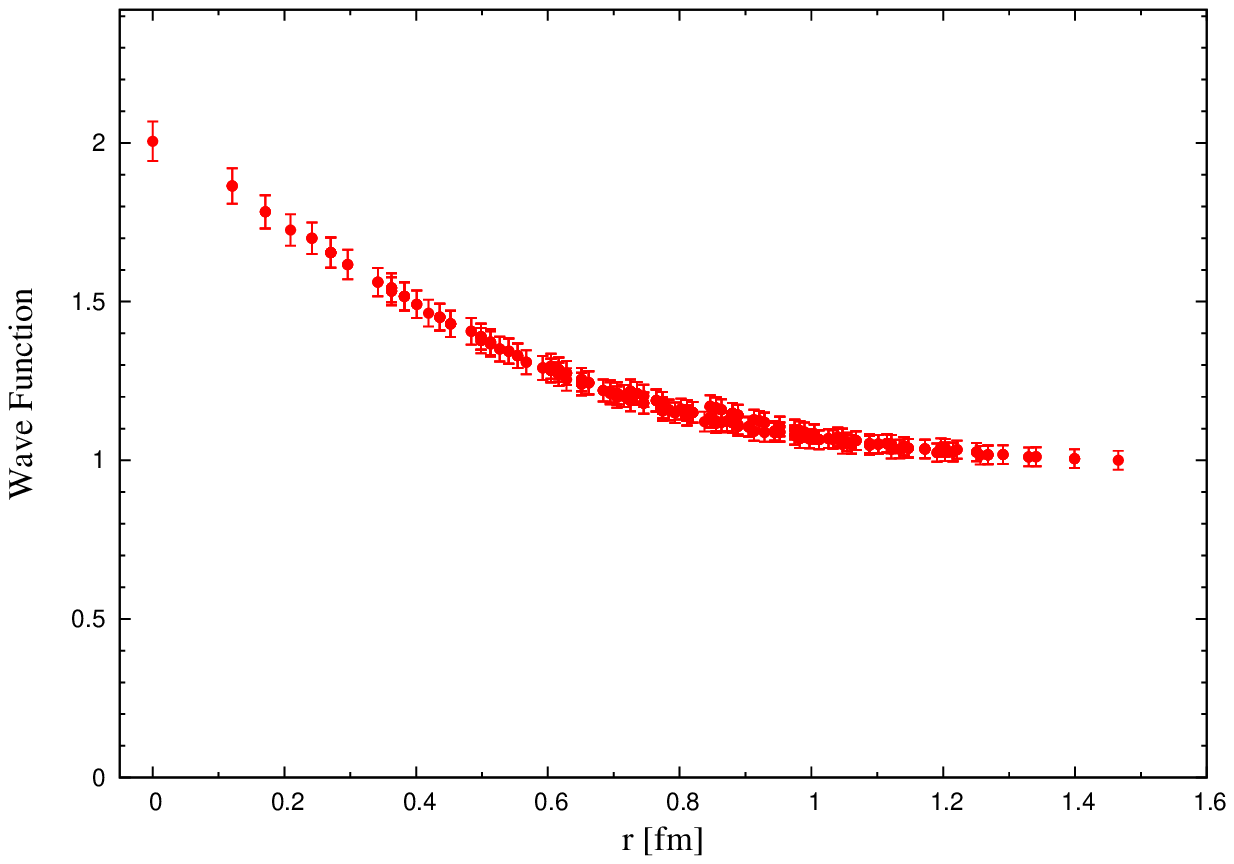}
\caption{The NBS wave function for $N\Omega$  in the $^{5}S_{2}$ channel
at $t-t_0=8$. The wave function is normalized to $1$ at the maximum distance 
by multiplying an overall factor.   Only statistical errors are shown.}
\label{fig:NBS}
\end{center}
\end{figure}
%--------------------
We employ 700  gauge configurations generated by CP-PACS and JLQCD
Collaborations\cite{cppacs-jlqcd}
 with the renormalization group improved
Iwasaki gauge action and the nonperturbatively $\mathcal{O}(a)$
improved Wilson quark action\cite{Ishikawa:2007nn} at
$\beta=1.83$ ($a\simeq 0.12$ fm ) on the $16^{3}\times32$ lattice.
    The Dirichlet (periodic)
boundary condition is imposed for quarks in the temporal (spatial)
direction. The wall source is used with the Coulomb gauge fixing, while
the sink operator is projected to the $A_{1}^{+}$ representation
of the cubic group, so that the NBS wave function is dominated
by the S-wave component. A number of sources per configuration is 8.
 Statistical errors in this report are estimated by the Jack-Knife method with a bin size of 100 configurations,
 though errors do not change much as long as the bin size is larger than 10 configurations.
 The hopping parameters  in our calculation are
 $(\kappa_{ud},\kappa_{s})=(0.13760,0.13710)$, and the
corresponding hadron masses  obtained with 32 sources was  given  in Table 4 (Set 1) of Ref.~\cite{HAL2012}:
$m_N=1806(3)$ MeV, $m_\Lambda = 1835(3)$ MeV and $m_\Xi=1867(2)$ MeV.
In addition,  we use  $m_\Omega=2105(5)$ MeV for our study, calculated from 8 sources with the fitting range $t-t_0=7-11$, whose effective mass is shown in Fig.\ref{fig:emass}.

The S-wave $N\Omega$ system with spin 2 can decay into the 
 D-wave $\Lambda\Xi$ system in the real world, 
 though such coupling is expected to be suppressed kinematically. 
 On our lattice setup with relatively heavy quarks, such a  D-wave threshold is located at 
$W_{\min}= \sqrt{m_\Lambda^2 +p_{\rm min}^2} + \sqrt{m_\Xi^2 +p_{\rm min}^2}\simeq$ 3920 MeV 
for  $p_{\rm min} =2\pi/L \simeq$ 645 MeV. This is 
 slightly above $m_N + m_\Omega \simeq 1806+2105=3911$ MeV, so that 
 we focus only on the $N\Omega$ channel in this report. 
More sophisticated analysis  using the coupled-channel HAL QCD method
 \cite{Aoki:2011gt,Aoki:2012bb,Aoki:2013cra,Sasaki:2013hta} 
 is left for future studies.

Let us  first show the $N\Omega$ NBS wave function in Fig.~\ref{fig:NBS}  
as a function of the relative distance $r$ in the $^{5}S_{2}$ 
channel at $t-t_0=8$.
The wave function is normalized to $1$ at the maximum distance by multiplying an
overall factor, which does not affect the potential and observables. 
The NBS wave function increases as the distance decreases, suggesting the attractive 
interaction between $N$ and $\Omega$.
%------------------
\begin{figure}[tbh]
\begin{center}
\includegraphics[scale=1]{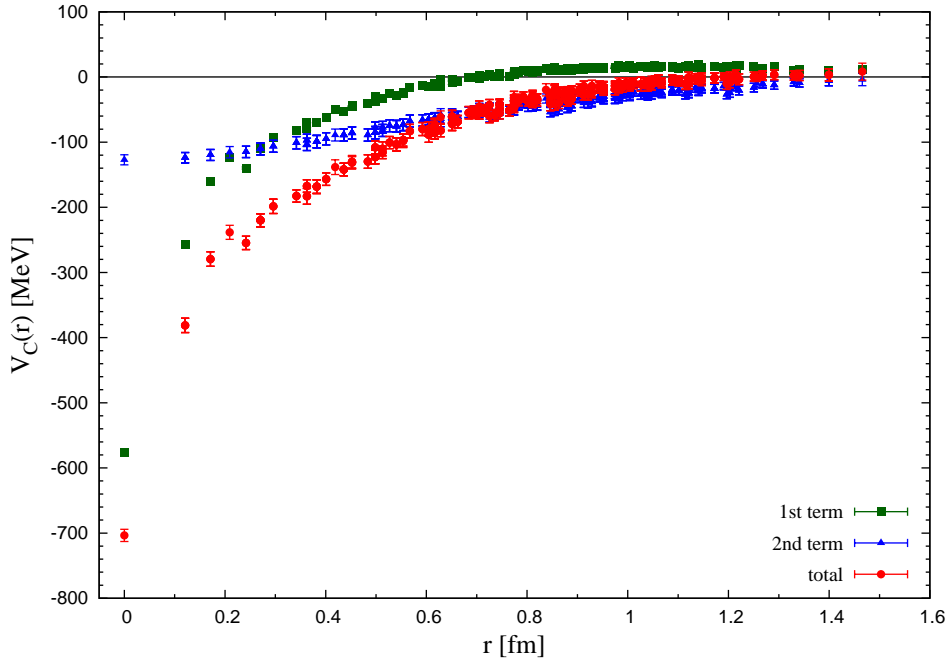}
\caption{The effective central potential $V_{C}(r)$ (red circles) for $N\Omega$ in the
$^{5}S_{2}$ at  $t-t_0=8$, together with its breakup, the first (green squares) and the second (blue triangles) terms in Eq.~(\ref{eq:t-dep}). Only statistical errors are shown. }
\label{fig:potential}
\end{center}
\end{figure}

\begin{figure}[tbh]
\begin{center}
\includegraphics[scale=1]{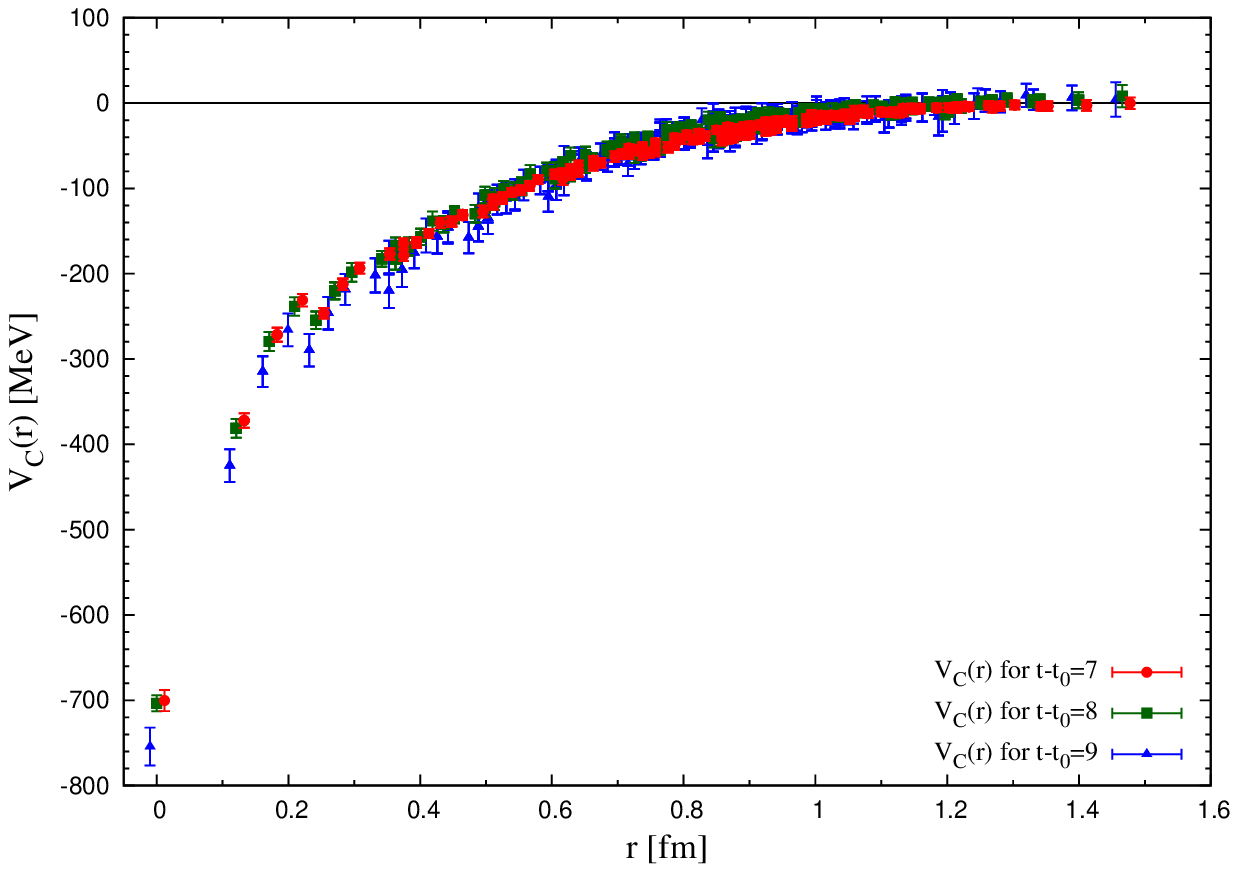}
\caption{A comparison of the effective central potential $V_{C}(r)$  for $N\Omega$ in the
$^{5}S_{2}$ at  $t-t_0=7$ (red circles), 8 (green  squares) and 9 (blue triangles).
For visibility, data are a little shifted horizontally at each $t-t_0$. Only statistical errors are shown.}
\label{fig:t-dep}
\end{center}
\end{figure}
%-----------------

In Fig.~\ref{fig:potential}, we 
show the $N\Omega$ effective central potential $V_{C}(r)$ in the 
$^{5}S_{2}$ channel at $t-t_0=8$; the first and the 
second terms in Eq.~(\ref{eq:t-dep}) are separately 
shown by green square and blue triangle points, respectively, 
together with the total potential $V_C(r)$ shown by the red circle points.
Note that the second term in Eq.~(\ref{eq:t-dep}) is essential to 
 extract potentials reliably from 4-pt functions even in the presence of excited 
 state contributions\cite{HALQCD:2012aa}.
 We find that  $V_C(r)$ in Fig.~\ref{fig:potential} is attractive at all distances
   as expected from the behavior of the NBS wave function in Fig.~\ref{fig:NBS}.

 Important observation emphasized in \cite{HALQCD:2012aa} is that
 the ground state saturation ($t \gg t_0$) is not necessary in our
  formulation and one can 
 extract the non-local potential by using the information from finite $t-t_0$ as long as 
  the inelastic threshold does not couple strongly. Furthermore, one can check
   the validity of the velocity expansion by looking at the $t$-dependence
   of the resultant local potentials. In Fig.~\ref{fig:t-dep},
  we compare $V_C(r)$ for $t-t_0=7\sim 9$: 
   Since they are consistent with each other within the error bars for all $r$,
    we conclude that the velocity expansion is working well for the low-energy scattering 
    between $N$ and $\Omega$.

\section{Scattering phase shift and binding energy}
\label{sec:observables}

\begin{figure}[tbh]
\begin{center}
\includegraphics[scale=1]{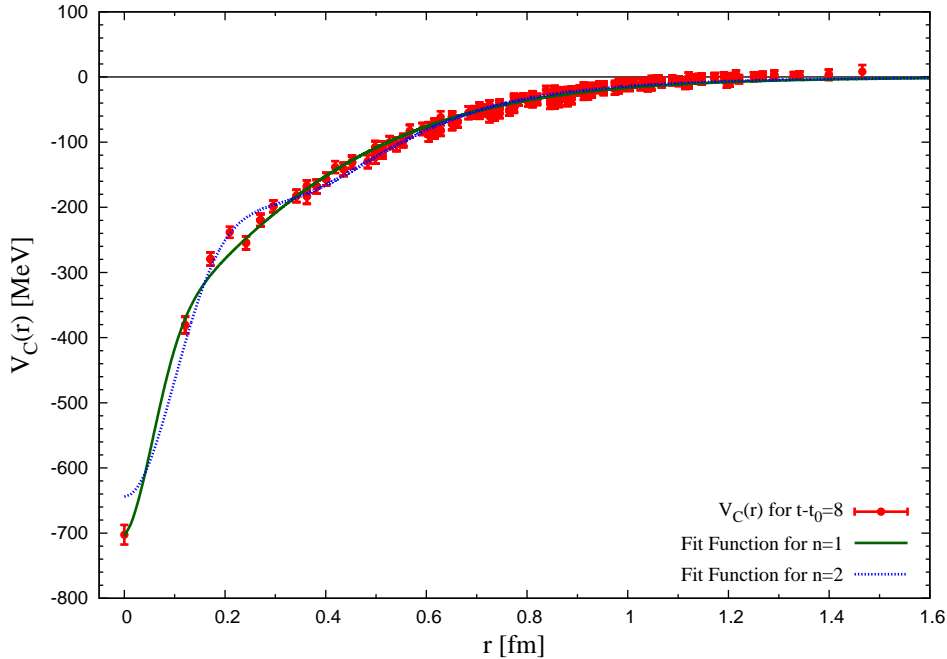}
\caption{Fit of the effective central potential $V_C(r)$ at $t-t_0=8$ for $n=1$ (solid dark-green line)
and for $n=2$ (dotted blue line). only statistical errors are shown.}
\label{fig:fit}
\end{center}
\end{figure}

In order to calculate the $N\Omega$ scattering phase shift from the potential obtained in the 
 previous section, we fit $V_C(r)$ with the Gaussian + (Yukawa)$^2$ form adopted in 
  our previous analysis of the H-dibaryon in lattice QCD\cite{Inoue:2010es}:
\begin{equation}
V_C(r)=b_{1}e^{-b_{2}r^{2}}+b_{3}(1-e^{-b_{4}r^{2}})^n(e^{-b_{5}r}/r)^{2},
\end{equation}
where we take either $n=2$ (the same form as \cite{Inoue:2010es}) or $n=1$.
Fitted results for the potential at $t-t_0=8$
are shown in Fig.~\ref{fig:fit} by a solid dark-green (dotted blue) line for $n=1$ ($n=2$) with
$\chi^2/$d.o.f. =0.95  (0.93).

%--------------------------------

%------------

\begin{figure}[tbh]
\begin{center}
\includegraphics[scale=1]{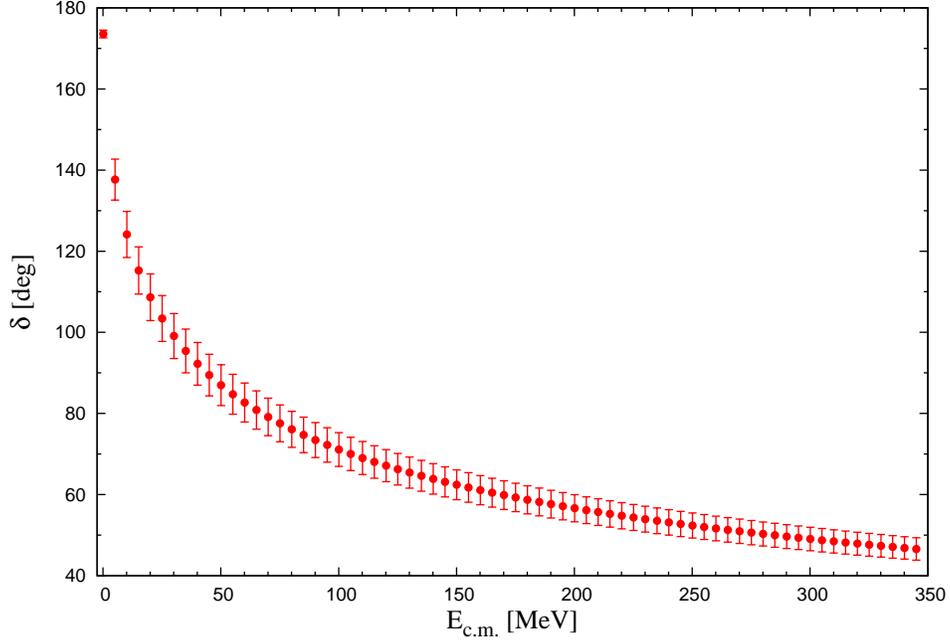}
\caption{Scattering phase shift $\delta$ as a function of the kinetic energy $E=k^2/(2\mu)$ in the center of mass frame, obtained 
from the potential   with the $n=1$ fit at $t-t_0=8$. Only statistical errors are shown.}
\label{fig:phase_shift}
\end{center}
\end{figure}

Using the fitted potential $V_C(r)$, 
we solve the
Schr\"{o}dinger equation in the infinite volume to calculate the $N\Omega$ scattering phase shift  
in the $^{5}S_{2}$ channel.
Fig.~\ref{fig:phase_shift} shows the scattering phase shift $\delta(E)$ extracted from the potential at
 $t-t_0=8$ with $n=1$,  as a function of the kinetic  energy $E=k^2/(2\mu)$ in the center of mass frame. (The results for $n = 2$ gives negligible difference from those for $n =1$ compared to the statistical errors.)  Note that systematic errors associated with the leading order approximation $V$  in  
 the velocity expansion for the non-local potential $U$ may become more sizable at larger kinetic energies. 

The scattering phase shift becomes 180 degrees
at $E=0$ and rapidly decreases as  $E$ increases,  which implies the 
 existence of a bound state in this channel. We thus calculate the binding energy $B$,
 which is given in Table~\ref{tab:results} at each $t-t_0$ with $n=1$ and 2, together with
the scattering length $a$ and the effective range $r_e$, 
 where the scattering length and the effective range are defined from the scattering phase shift $\delta(E)$ as
 \begin{eqnarray}
k \cot\delta(E) &=&\frac{1}{a} +\frac{r_e}{2} k^2 +O(k^4) .
\end{eqnarray}

 \begin{table}[hbt]
 \caption{The binding energy, the scattering length and the effective range obtained from each  potential at $t-t_0$ with $n=1,2$.}
 \label{tab:results}
 \begin{center}
  \begin{tabular}{|c|c|ccc|}
    \hline
    $n$ & $t-t_0$  & $ B$ (MeV)   & $a$ (fm)   & $r_e$ (fm)  \\
    \hline
     1 & 6 & 17.9 (2.8) & -1.57 (0.10) & 0.532 (0.018)  \\
        & 7 & 16.2 (3.6) & -1.63 (0.15) & 0.538 (0.023) \\
        & 8 & 14.6 (5.8) & -1.69 (0.26) & 0.537 (0.035) \\
        & 9 & 24.5 (11.5) & -1.35 (0.28) & 0.468 (0.049) \\
        & 10 & 46.9 (20.7) & -1.04 (0.20) & 0.407 (0.075) \\
        & 11 & 38.3 (31.4)  & -1.18 (0.35) & 0.473 (0.142) \\
     \hline
     2 & 6 &  18.3 (3.0) &  -1.43 (0.15) & 0.532 (0.022)    \\
         & 7 &  16.4 (3.9) & -1.53 (0.19) & 0.541 (0.031) \\
         & 8 & 15.3 (6.0) & -1.57 (0.23)  & 0.556 (0.044) \\
         & 9 & 25.0 (11.7) & -1.27 (0.29)  & 0.497 (0.059) \\
         &10 & 46.3 (20.7) & -1.08 (0.17) & 0.412 (0.082) \\
         &11 & 39.1 (31.4) & -1.18 (0.33) & 0.468 (0.135) \\
    \hline
  \end{tabular}
 \end{center}
\end{table}

Since effective masses of both $N$ and $\Omega$ show plateau  for $t-t_0 \ge 7$,
 the central value and statistical error of the observables are estimated from the weighted average over  
 $8 \le t-t_0 \le 11$ with $n=1$.  To estimate  the systematic errors, 
 we consider the observables obtained from the average over   $7 \le t-t_0 \le 11$ and  over $9 \le t-t_0 \le 11$,
 as well as those from the average over $t_{\rm min}\le t-t_0\le 11$ $(t_{\rm min}=7,8,9)$ with $n=2$.
Finally we  obtain
\begin{eqnarray}
B_{N\Omega}&=&18.9(5.0)(^{+12.1}_{-1.8})\  {\rm MeV}, \label{eq:B}\\
a_{N\Omega} &=& -1.28 (0.13)(^{+0.14}_{-0.15}) \  {\rm fm},\label{eq:a}\\
(r_e)_{N\Omega} &=& 0.499 (0.026)(^{+0.029}_{-0.048}) \ {\rm  fm}. \label{eq:r}
\end{eqnarray}
 Here the numbers in the first parenthesis correspond to the statistical error, while 
 those in the second parenthesis show the systematic errors
 obtained by taking the largest difference between the central value and the other 5 values.
  Note that this systematic uncertainty is still sizable, in particular for the binding energy.

\section{Summary and concluding remark}
\label{sec:conclusion}

We study the $N\Omega$ interaction in the $^{5}S_{2}$ channel through
the equal-time NBS wave function in (2+1)-flavor lattice QCD simulations 
on a (1.92fm)$^{3}\times 3.84$ fm lattice at  the quark masses corresponding to $m_{\pi}=875(1)$ MeV
and   $m_{K}=916(1)$ MeV.

We extract the central potential from the NBS wave function for the $^5S_2$ channel by using the
 time-dependent HAL QCD method. The result shows attraction for all distances in this channel.
We found that the attraction is strong enough to produce one $N\Omega$ bound state in the spin-2 channel
at our quark mass.
In our study we  obtain 
$B_{N\Omega}=18.9(5.0)(^{+12.1}_{-1.8})$  MeV,
$a_{N\Omega} = -1.28 (0.13)(^{+0.14}_{-0.15})$ fm and  
$(r_e)_{N\Omega} = 0.499 (0.026)(^{+0.029}_{-0.048})$ fm.
  
 Our present study may be considered as a starting point to answer 
 the long standing question about the existence of the 
  $N\Omega$ bound state in the spin-2 channel. We are currently studying 
 the  $N\Omega$ potential by using gauge configurations at larger lattice volume with smaller quark masses
 generated by PACS-CS collaboration\cite{Aoki:2008sm}.  
 In the near future, large volume simulations  at the physical quark masses
 using  K-computer at RIKEN AICS \cite{Aoki:2012oma}  with the coupled-channel HAL QCD method would make 
 a final conclusion on the fate of the $N\Omega$ bound state.

\section*{Acknowledgment}

We thank CP-PACS/JLQCD  Collaborations and ILDG/JLDG for providing us the
$2+1$ flavor gauge configurations. 
We are grateful for the authors and maintainers
of Bridge++\cite{bridge}, a modified version of which is used for calculations
in this work.
T.H. would like to express his sincere thanks to the late Prof. Gerry Brown 
who gave him kind and constant encouragement for many years.
F.E. thanks Prof. M. M. Firoozabadi for his support. He is also supported in part by 
Department of Physics Bilateral International Exchange Program (BIEP) 2013, Kyoto University, and thanks 
Yukawa Institute for Theoretical Physics, Kyoto University for a kind hospitality during his stay while completing this paper.
This research is supported in part by Grant-in-Aid
for Scientific Research on Innovative Areas(No.2004:20105001, 20105003) and
for Scientific Research (B) 25287046, 24740146, (C) 23540321 and SPIRE (Strategic Program for Innovative Research). T.H. is partially supported by RIKEN iTHES project.

%\clearpage

\end{document}